# Exploring the van der Waals Atom-Surface attraction in the nanometric range

M. Fichet, G. Dutier, A.Yarovitsky[§], P. Todorov, I. Hamdi, I. Maurin,

S. Saltiel[*], D. Sarkisyan[#], M.-P. Gorza, D. Bloch and M. Ducloy

*Laboratoire de Physique des Lasers, UMR7538 du CNRS et de l'Université Paris13.*

*99, Av. J.B. Clément, F-93430 Villetaneuse, France*

[§]*Permanent address : Lebedev Physics Institute, Moscow*

[*]*Also at Physics Department, Sofia University, Bulgaria*

[#] *Institute for Physical Research, National Academy of Sciences, Ashtarak 2, Armenia*

*correspondence to be addressed to: bloch@lpl.univ-paris13.fr*

         34.50.Dy        Interactions of atoms and molecules with surfaces;
                         photon and electron emission; neutralization of ions
         32.70.Jz        Line shapes, widths, and shifts

*The van der Waals atom-surface attraction, scaling as $C_3 z^{-3}$ for z the atom-surface distance, is expected to be valid in the distance range 1-1000 nm, covering 8-10 orders of magnitudes in the interaction energy. A Cs vapour nanocell allows us to analyze the spectroscopic modifications induced by the atom-surface attraction on the $6P_{3/2} \rightarrow 6D_{5/2}$ transition. The measured $C_3$ value is found to be independent of the thickness in the explored range 40-130 nm, and is in agreement with an elementary theoretical prediction. We also discuss the specific interest of exploring short distances and large interaction energy.*

The dipole-induced dipole attraction between neutral bodies is a key phenomenon in the ultimate cohesion of matter and is at the origin of covalent crystals and van der Waals molecules. This van der Waals (vW) type of attraction between fluctuating atomic dipoles is a precursor of the general Casimir interaction [1], whose paramount importance in the quantum theory of vacuum has been recognized recently, encompassing effects related to the need of a "cosmological constant" in general relativity theory, and speculations about the detection of a possible fifth force [2]. The interest for the measurement of the Casimir effect has been largely renewed with the recent upsurge of precision experiments [1,3], whose interpretation now requires to take into account various corrections such as the non-ideality of the materials, surface roughness, and non-zero temperature corrections. Simultaneously, the knowledge of the vW attraction between atomic particles, with its $-C_6 r^{-6}$ potential (with r the interatomic distance), now appears as a fundamental need to predict the collisional behavior of a collection of atoms [4], ultimately connected to the possibility for cold atoms to condense according to a Bose-Einstein statistics. Between these two related fundamental problems, an intermediate situation is provided by the atom-surface vW interaction, described by a $-C_3 z^{-3}$ potential (z: the atom-surface distance). To theoretically estimate the $C_3$ value from the knowledge of atomic structures, an electrostatic mirror image approach is usually satisfactory in many cases. The $z^{-3}$ dependence is expected to be valid for distances allowing to ignore the structural details of the surface (typically $\geq 1$ nm), up to the micrometric range, when retardation effects impose the more complete Casimir-Polder description [2,5].

It remains fascinating to note that, although the vW $z^{-3}$ attraction law should cover about 8-10 orders of magnitude of energy, little has been done to test this predicted dependence in an effective situation. The various developments in nanotechnologies and nanosciences should make it more important to measure effectively these remote forces, as for the Casimir force already known to be a possible limiting factor in MEMS technology [6]. In the shorter distance limit, the vW attraction is only an asymptotic "long-range" tail of the atom-surface potential, to the exact shape of which surface physics experiments are insensitive [7]. For longer distances, and after pioneering principle experiments using deflection of an atomic beam [8], precision experiments tested the spatial dependence on a beam of Rydberg atoms [9] channelled between two plates separated by at least 500 nm. Following the blossom of experimental measurements of vW interaction exerted on a long-lived level [10,11], the spatial dependence of vW interaction between a ground state Rb atom and a metal surface was recently investigated in the 20-100 nm range [12]. Selective Reflection (SR) spectroscopy of a vapour at an interface, in which we have been involved for a long time [13], has offered a unique capability to probe the vW interaction for short-lived excited states. It is also appropriate for analyzing the atom interaction with a (transparent) dielectric surface. Although a dielectric surface commonly exerts a weaker interaction than a metal, owing to an imperfect reflection, a specific regime of resonant coupling of the atomic excitation with the electromagnetic modes of the dielectric surface, could be demonstrated with this SR technique, leading to an exotic vW repulsion [14]. However, the depth of the region probed in the SR technique remains fixed, as governed by the (reduced) wavelength of the optical probe [13]. Here, we use an alternate technique of spectroscopy in a vapour "nanocell" [15], whose thickness d is varied in a range d = 40-130 nm. This allows to explore the spatial dependence of the vW interaction.

For a *beam* of atoms flying parallel to the two windows and at mid-distance (see e.g. [9]) a constant vW shift of the atomic resonance, scaling as $1/d^3$, is expected, even if multiple (dielectric) images are considered (the vW shift is here a *spectroscopic* shift, corresponding to the *difference* between the vW interaction exerted onto the final level of the transition, and the one exerted onto the initial one). However, for atoms in a *vapour* phase, the vW shift is spatially inhomogeneous, z = d/2 is only the location of the minimal vW shift, and the lineshapes undergo a spatial inhomogeneous broadening leading to distortions and asymmetries. In this sense, the preliminary observation of an elementary $1/d^3$ dependence of the frequency of the minimal transmission [17] was not a sufficient evidence of the vW dependence. The principle of our estimate of the vW interaction for a given nanocell thickness d relies on a numerical comparison between the experimental spectra and a family of relevant theoretical lineshapes.

Our experiment was conducted on the $6P_{3/2}$ - $6D_{5/2}$ transition of Cs ($\lambda$=917 nm) in a nanocell with YAG windows. The choice of a transition between excited states is to allow a strong vW shift, mainly originating in the vW interaction exerted onto the high-lying 6D level (fig.1) [14] : one predicts indeed $C_3^{metal}(6D_{5/2})$ = 24 kHz.µm$^3$ >> $C_3^{metal}(6P_{3/2})$ = 4 kHz.µm$^3$, and in front of a YAG window: $C_3^{YAG}(6D_{5/2})$ = 17 kHz.µm$^3$, $C_3^{YAG}(6P_{3/2})$ = 2 kHz.µm$^3$, yielding a *spectroscopic* $C_3$ value of ~15 kHz.µm$^3$ for YAG windows. To reach the transition of interest, a prior pumping step to Cs($6P_{3/2}$) is needed, which is performed with a 852 nm frequency-stabilized laser (see fig.1). YAG windows were preferred to sapphire in the design of the nanocell because in the atom-dielectric surface, there are no strong couplings between YAG surface resonances and virtual transitions relevant for Cs($6D_{5/2}$) [14]. The nanocell, once evacuated, is filled with an alkali-metal vapour, and consists of a system of two thick windows, that are glued and maintained separated by a nanometric spacer [15]. The external atmospheric pressure is responsible for local variations of the nanocell thickness [15], so that translating the cell through a focused laser allows to monitor the influence of the thickness under unchanged experimental conditions. As already reported, the Fabry-Perot nature of the nanocell allows a convenient determination of the local thickness d [15]. Here, for the very small thicknesses we are interested in, a dual-wavelength measurement (633 nm and 852 nm) is sufficient: the accuracy is 1 to 5nm, depending upon the parallelism of the windows at the considered spot, whose diameter is ~100 µm. Also, the Fabry-Perot nature of the nanocell allows the detection of both transmission and reflection signals (respectively $S_T$, and $S_R$) [15, 16]. The spectra were simultaneously recorded, for thicknesses d = 40 nm, 50 nm, 65 nm, 80 nm, 100 nm and 130 nm. The detection sensitivity is enhanced by applying a modulation to the $\lambda$=852 nm pumping beam, with a subsequent synchronous detection on the $\lambda$=917 nm probe. The transition in free-space is monitored by an auxiliary saturated absorption (SA) experiment on the 917 nm line in a macroscopic cell and provides the frequency reference required for interpreting the nanocell spectra. In an approach analogous to the one developed previously (*e.g.* in [14]), experiments were performed at several Cs densities -*i.e.* temperature- for each investigated spot of the nanocell. This is to minimize the influence of Cs pressure on the extracted $C_3$ value. Also, a variety of pumping conditions was explored to assess the spatial homogeneity of the $6P_{3/2}$ population.

In the theoretical modelling, the transmission and reflection signals $S_T$ and $S_R$ are derived from linearly independent integrations [16] of the local vapour response p(z) as governed by the resonantly induced atomic dipole oscillation. This atomic oscillation is modelled on the basis of an instantaneous resonant Lorentzian response $\{\gamma/2 +i[\omega-\omega_0(z)]\}^{-1}$, that depends upon the detuning between the excitation frequency $\omega$ and the local vW-shifted atomic resonance $\omega_0(z)$ ($\gamma$ is the optical width). It exhibits a nonlocal dependence upon the irradiating field, owing to the transient regime governing the (velocity-dependent) atomic response over the wall-to-wall trajectories. In the modelling, the unshifted transition ($\gamma$, $\omega$) can possibly accomodate pressure broadening and shift. Hence, $S_T$ and $S_R$ integrate a complex interplay of natural width, Doppler broadening and velocity distribution, pressure effects, and spatial dependence of the vW potential. Our complete calculation even considers multiple electrostatic images for the vW potential -although a two-wall potential $V(z) = - C_3[z^{-3} + (d-z)^{-3}]$ agrees within 5 % as long as the dispersion of the dielectric window [14] is negligible-. Strong variations of the lineshapes are hence predicted when changing d, or $C_3$ [18].

However, when the vW interaction dominates over other causes of broadening, the spectral shift follows a $1/d^3$ dependence rather well. For parameters relevant to our problem, the modelling shows that the 40-130 nm range is sufficient to explore a factor $\geq 30$ in the energy of surface interaction (see fig. 2).

Technically, the analysis relies on a fitting of the experimental spectra $S_T(\omega)$ and $S_R(\omega)$ with the family of $C_3$ curves calculated for a given thickness d. The strong constraints on the vW shift and lineshape restrict the $C_3$ values providing a satisfactory (least-square) fitting to a narrow-range, typically 10-20% around the central value. When the vW shift is dominant, the vW spatial broadening makes fittings nearly insensitive to the adjustment of the width of the optical transition. The main results of our systematic analysis are two-fold : (i) for a given experiment, a very good consistency between the shapes of $S_T$ and $S_R$ signals is demonstrated [18] ; their relative amplitudes are also in agreement once considered the scattering losses [19]; this consistency is remarkable because it appears for lineshapes that are unrelated, and with fitting parameters that are independently chosen; (ii) the optimal fitting $C_3$ values are found to be independent of the thickness. Figure 3 illustrates synthetically this consistency: all spectra recorded in identical conditions, but for various thicknesses, are fitted with a single set of parameters. The residual discrepancies in the fitting disappear with more individual adjustments, and the individual $C_3$ fitting values are modified only marginally. Fig.4 is a plot of the individually optimized fitting value for $C_3$ for differing thicknesses, and pressure conditions. When retaining only the experiments at relatively low pressure, one gets an accurate value $C_3 \approx 14 \pm 3$ kHz.µm$^3$. For the smallest thickness (*i.e.* dominant vW contribution), the obtained $C_3$ values are independent on the Cs pressure, in spite of a pressure broadening visible in the lineshapes. For larger thickness ($d \geq 80$nm), the vW shift and broadening become partly hindered by pressure effects, and decreasing $C_3$ values are found when increasing the Cs density. And indeed, for the large thicknesses, when the vW interaction turns to be relatively small, the accuracy of the $C_3$ determination starts to be affected by uncertainties on the 917 nm SA reference frequency itself: an uncontrolled preferential pumping to a specific $6P_{3/2}$ hyperfine sublevel is susceptible to occur, owing to nonlinearities in the $D_2$ pumping.

Notwithstanding the excellent agreement between the experimental data and the theoretically modelled curves, the reproducibility issues are worth being discussed, as well as the effective nature of the interacting surface. Reproducibility in the vicinity of a dielectric interface is shown to be an issue in [2], with numerous erratic points obtained, even with well-chosen surfaces. In our experiment, the measured thickness is an average, over the laser-spot (diameter ~ 100 µm) of local wall-to-wall distance d. Because of the local surface roughness (estimated to 1-2 nm), and of the defects in planarity or parallelism (the minimal observed slope of the nanocell can be as small as ~ 10 nm for 1mm [15]), the observed vW effect can vary for comparable spots as $<d>^{-3} \neq <d^{-3}>$. The surface of our nanocell largely exceeding the spot size, several experiments with a similar nominal (average) thickness can be compared. A perfect lineshape reproducibility has been found for all the various investigated spots, as long as $d \geq 65$ nm. At $d = 50$ nm, several spots -but not all of them- produce rigorously identical spectra: they are those here analyzed, as apparently immune to random defects of the surface that would not be compatible with the observation of similar spectra on a differing spot. For $d = 40$ nm, sensitive variations are observed from spot to spot; however, except for a few erratic behaviours clearly out of the considered family of lineshapes, minor irreproducibilities do not prohibit a fitting as exemplified in fig.3. Remarkably, in this regime of very strong vW interaction, minor changes in the fitting parameters suffice to interpret visible variations in the lineshapes.

The exact nature of the interacting surface can also be an issue, especially because the surface is in contact with a vapour, and because for a dielectric surface, relatively high static electric fields can induce Stark shifts. In a vapour, Cs atoms are susceptible to deposit on, or penetrate the YAG surface; however, the dielectric function characterizing the vW surface response should remain nearly unaffected by these kinds of (non resonant) dopants [20]. The competition between the vW shift, and the Stark shift induced by a residual static field can pose a more serious problem, as recognized in [9]. Moreover, stronger residual fields can be expected for a dielectric window [21] than for a metal coating. In particular, for some random distributions of crystalline domains, the overall Stark shift may exhibit a $z^{-2}$ dependence [9], that could be hard to distinguish from a vW shift in the spectroscopic response (conversely, a constant Stark shift would be easily detected in an analysis of our fittings). This is a severe problem for experiments on Rydberg levels, because when increasing the atomic excitation, the Stark shift grows faster than the vW shift (respectively like $n^{*7}$, and $n^{*4}$ in common approximations, for n* the effective quantum number). From our own experience with Cs* close to a YAG window, we can conclude that the influence of residual fields should remain minor: a SR experiment on the Cs(7P→10 D) line at 1.3 µm [22] (at a comparable YAG interface) had indeed shown the Stark shift to be negligible (i.e. ≤ 100 MHz). Extrapolation from Cs(10D) (n*≈7.5) to Cs(6D) (with n*≈3.5) hence implies a maximal Stark shift on the order of ~MHz in similar conditions. Even if a possible $z^{-2}$ dependence is considered, the influence of the Stark shift should remain below several 100 MHz for the smaller nanocell thicknesses, and be even less for larger thicknesses. This is why we estimate that the major causes of irreproducibility are connected to uncertainties, at the level of a few nm, in the local geometry of the two planar windows, rather than to the effect of stray fields.

Our estimated value ($C_3$ =14 kHz.µm$^3$) is in very good agreement with the ~15 kHz.µm$^3$ theoretical prediction. This agreement may be considered as remarkable in view of various pitfalls, either experimental, or theoretical. A precise evaluation of the dielectric image factor applicable to Cs(6D$_{5/2}$) is delicate [23] because the situation is not purely nonresonant: the virtual *emission* coupling Cs($6D_{5/2}$) to 7D (in the ~15μm region) requires the accurate knowledge of the YAG spectrum and of its surface modes. Moreover, in a nanocell [24], the atom couples to a guided mode structure, and the two-wall model, or its straightforward extension to multiple electrostatic images model, may reveal too elementary. Also, the orientation of the atom has been assumed to be isotropic, implying equal contributions for the dipole fluctuating along the normal, and parallel to the interface. However, the atom could undergo a Zeeman polarization under the influence of the $D_2$ line pumping (the irradiation, being under near normal incidence, is polarized *parallel* to the surfaces), or through the polarized excitation of the 917 nm laser. This would restrict the summation over the virtual dipole transitions connecting the $6D_{5/2}$ level to only some directions, reducing the predicted $C_3$ value [24]. More generally, a recent work [25] shows that $C_3$ itself should exhibit a spatial dependence $C_3(z)$, because at the smaller distances, the (non retarded) contribution of the highly energetic transitions involving the electronic core becomes stronger [4]. This dependence, actually unobserved [10,12], is however expected to be smaller in our situation of a high-lying state [13], than for a ground state.

To summarize, we have investigated the $z^{-3}$ dependence of the vW potential for a 40-130 nm thickness range, and an energy shift spanning over a factor ~30. Our technique could be extended to test the rich physics of an atom resonantly interacting with coupled surface modes. The distances investigated here are an order of magnitude below those explored years ago for excited atoms in Rydberg levels [9], and compares favourably with those currently investigated in precision Casimir measurements [3], or with the one addressed in the recent vW interaction experiments with ground state atoms [12]. Note that in our situation, the vW shift, whose minimal value is twice the one undergone by an atom at a distance d/2, is equivalent the one of an atom located at ~0.4d from a single wall. Also, the combined effects of the transient regime, and the steepness of the vW potential, make the central region of the nanocell spectrally dominant, as can be inferred from fig.2: hence, the individual evaluations $C_3(d)$ turn out to be nearly free of spatial averaging. Lowering the effective atom-surface distance below 10 nm appears a realistic objective (feasibility of d ~20-30 nm is reported in [17]). This is in contrast with the vW measurements based upon the reflection of slow atoms [10,12]: the minimal distance of approach of the observed atomic trajectories is limited by the considerable force that should be applied to equilibrate the vW attraction. Similarly, the techniques of atomic force microscopy presently used for the evaluation of Casimir interaction, are hardly compatible with too high a pressure: the standard calculation predicts indeed a Casimir interaction exceeding the atmospheric pressure for d ≤ 10 nm. Note that for these small distances, the asymptotic regime of van der Waals Casimir-related interaction remains under debate [26]. Finally, it is worth mentioning that in the small distance regime that we explore here, we demonstrate an interaction energy (up to ~5 GHz, or 0.25K) much larger than obtained in all previous investigations. This corresponds to a considerable acceleration (~ $8.10^7$ g for a Cs(6D) atom 20 nm away from one of the wall), exceeding by orders of magnitude the one obtained in laser cooling techniques. This may open a realm of exotic possibilities, such as a gradient of density for an atomic gas in the extreme vicinity with the surface.

*Work partially supported by FASTNet (European contract HPRN-CT-2002-00304) and by French-Bulgarian cooperation RILA (#09813UK ).*

**References**

[1] H.B.G. Casimir, *Proc. Kon. Ned. Akad. Wetenshap* **60**, 793 (1948); for reviews, see *e.g.* : M. Bordag, U. Mohideen and V. Mostepanenko, Phys. Rep. **353**, 1 (2001); A. Lambrecht and S. Reynaud, in *Poincaré seminar 2002*, *Vacuum energy*, (B.V. Rivasseau ed., Birkhauser,) p.109 (2003); K.A. Milton, J. Phys. A, **37**, R209 (2004).

[2] D. M. Harber *et al.*, Phys. Rev. A **72**, 033610 (2005)

[3] U. Mohideen and A. Roy, Phys. Rev. Lett. **81**, 4549 (1998); T. Ederth, Phys. Rev. A **62** 062104 (2000), F. Chen *et al.*, Phys. Rev. A **69**, 022117 (2004)

[4] A. Derevianko *et al.*, Phys. Rev. Lett. **82**, 3589 (1999).

[5] H.B.G. Casimir and D. Polder, Phys. Rev. **73**, 360 (1948)

[6] H.B. Chan *et al.*, Phys. Rev. Lett. **87**, 211801 (2001); E. Buks and M.L. Roukes, Europhys. Lett. **54** 220 ( 2001)

[7] F.O. Goodman and H.Y. Wachman, "Dynamics of Gas-Surface Scattering", Academic Press, 1976

[8] D. Raskin and P. Kusch, Phys. Rev. **179**, 712 (1969).

[9] V. Sandoghdar *et al.,* Phys. Rev . Lett. **68** 3432 (1992); Phys. Rev. A **53**, 1919 (1996).

[10] A. Landragin *et al*., Phys. Rev. Lett. **77**, 1464 (1996) ;

[11] R. E. Grisenti *et al*., Phys. Rev. Lett. **83**, 1755 (1999) ; M. Boustimi *et al*., Phys. Rev. Lett. **86**, 2766 (2001); F. Shimizu, Phys. Rev. Lett., **86**, 987–990 (2001).

[12] A.K. Mohapatra and C.S. Unnikrishnan, Europhys.Lett. **73**, 839 (2006)

[13] For a review, see D. Bloch and M. Ducloy "*Atom-wall interaction",* Adv. At. Mol. Opt. Phys., **50** pp. 91-154 (B. Bederson and H. Walther eds., Elsevier-Academic Press, 2005).

[14] H. Failache *et al.,* Phys. Rev. Lett. **83**, 5467 (1999); Eur. Phys. J. D, **23**, 237 (2003).

[15] D. Sarkisyan *et al*., Opt. Commun. **200**, 201 (2001); G. Dutier *et al*., Europhys. Lett. **63**, 35 (2003)

[16] G. Dutier *et al*., J. Opt. Soc. Am. B, **20**, 793 (2003).

[17] G. Dutier *et al*., in "*Laser Spectroscopy, Proceedings of the XVI International Conference*", (P. Hannaford *et al*., eds., World Scientific, Singapore, 2004) pp.277. These measurements do not discriminate between atom-surface interaction and collision processes; see also I. Hamdi *et al*., Laser. Phys, **15**, 987 (2005).

[18] I. Maurin *et al*., J. Phys. Conf Ser. **19**, 20 (2005). In this reference, the $C_3$ values given are actually $r_p C_3$, with $r_p$ (=0.536) the dielectric image factor, because of an accidental mistake in the transcription between computer values and the meaning of the fitting parameter.

[19] Experimentally, reflection coefficients on the windows do not accurately comply with the Fresnel formulae because of scattering. This explains variations from spot to spot in the ratios between $S_T$ and $S_R$.

[20] M. Fichet *et al*, Phys. Rev. A **51**, 1553 (1995)

[21] Static charges and random distribution of crystalline domains have been apparently responsible for erratic signals in an experiment derived from ref.9, but with the atomic beam of Rydberg atoms passing between dielectric walls (S. Haroche, private communication).

[22] H. Failache *et al*., Phys. Rev. Lett. **88**, 243603 (2002)

[23] S. Saltiel, D. Bloch and M. Ducloy, Opt. Commun. **265**,220-233 (2006)

[24] M.-P. Gorza, in preparation

[25] A. O. Caride *et al*., Phys. Rev. A **71**, 042901 (2005).

[26] see *e.g.* M. Scandurra, arXiv:hep-th/0306076/v2 (2003) ; Note that for the early experiments of van der Waals interaction between two surfaces performed in the nanometric range of distances, such as in J.N. Israelachvili and D.Tabor, Proc. Roy. Soc. A, **331**, 19-38 (1972), the geometry actually averages on distances much larger than the minimal distance, whose estimate results from an optical measurement integrating much larger distances.

**Figure captions**

Figure 1 : Scheme of the relevant energy levels of Cs and of the experimental set-up. $S_R$ and (inverted) $S_T$ are provided by processing by the reflected and transmitted intensities through lock-in detectors.

Figure 2: vW-induced spectral shift between the dip of minimal transmission and the free-atom resonance as a function of the thickness. The calculation is performed for: $C_3 = 14$ kHz.µm$^3$ , $\gamma = 300$ MHz.
The dotted line is for a gas of atoms flying wall-to-wall with a 250 MHz Doppler width. The dashed line, and the continuous line, are respectively for a beam of atoms flying at mid-distance d/2, and for a gas of motionless atoms.

Figure 3: Experimental lineshapes ($S_R$ and inverted $S_T$) recorded on the 917 nm transition for various thicknesses. The frequency scans are continuous, or discreet (for 40 nm and 50 nm). The vertical dashed line is a marker for the SA resonance. All fittings (dotted lines) use the parameters optimized for $S_T$ (50 nm), found to be $C_3 = 14$ kHz.µm$^3$, $\gamma = 300$ MHz. Adjustable amplitudes are introduced to compensate for the thickness dependence of the efficiency of the $6P_{3/2}$ pumping. For 40 nm: respectively 3.3 and 4.8 for $S_T$ and $S_R$; for 65 nm : 0.25 and 0.24; for 80 nm: 0.19 and 0.15; for 100 nm: 0.11 and 0.07; for 130 nm: 0.07 and 0.04. The Cs nanocell temperature is 200°C. The pumping frequency is locked onto the 4 → 4-5 crossover of the $D_2$ line, pump power is ~1 mW focused on a diameter ~ 100 µm, pump absorption reaches 25 % for 130 nm. The 917 nm transmission change is ~ 5.10$^{-4}$ for 50 nm.

Figure 4: Optimal fitting $C_3$ values found for various thickness and Cs temperatures. Increasing the Cs temperature by 20°C approximately doubles the Cs density.

M.Fichet *et al*, Figure 1

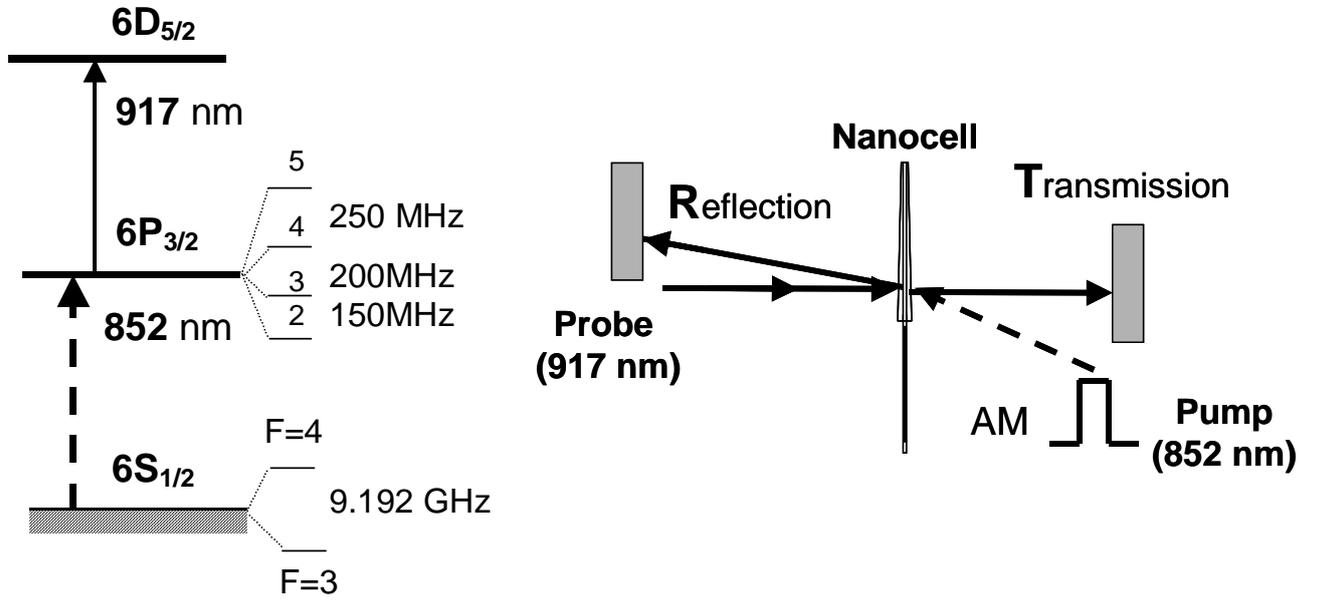

M.Fichet *et al*, Figure 2

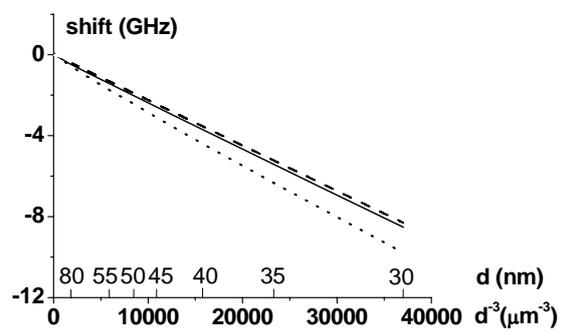

M.Fichet *et al*, Figure 3

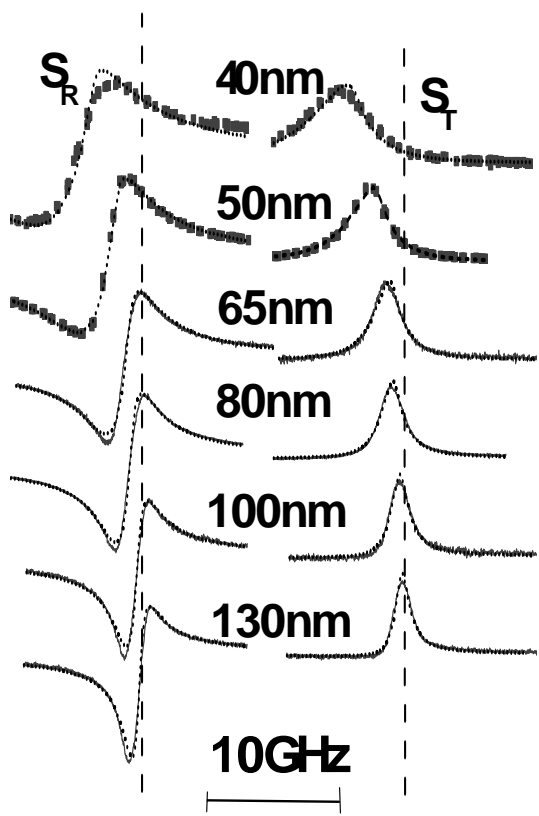

M.Fichet *et al*, Figure 4

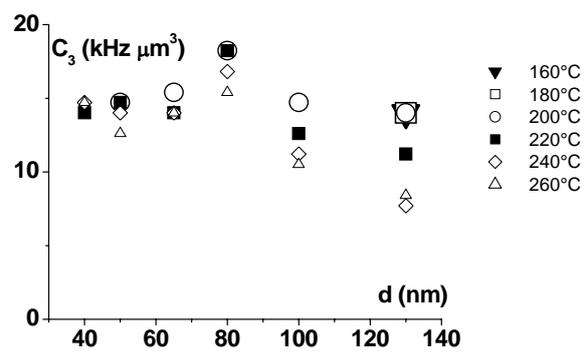